\documentclass[useAMS,usegraphicx,usenatbib]{mn2e}

\usepackage{graphicx}
\usepackage{mathptmx}
\usepackage{amssymb}
\usepackage{amsmath}
\usepackage{euscript}

\def\asca{{\it ASCA}}

\def\chandra{{\it Chandra}}

\def\xmm{{\it XMM-Newton}}

\def\rosat{{\it ROSAT}}

\def\s{{\rm\thinspace s}}
\def\ks{{\rm\thinspace ks}}

\def\gyr{{\rm\thinspace Gyr}}

\def\ctps{{\rm\thinspace ct\ s^{-1}}} 

\def\kev{{\rm\thinspace keV}}

\def\km{{\rm\thinspace km}}

\def\cm{{\rm\thinspace cm}}

\def\kpc{{\rm\thinspace kpc}}
\def\mpc{{\rm\thinspace Mpc}}

\def\kmps{{\rm\thinspace km s^{-1}}}

\def\g{{\rm\thinspace g}}

\def\cmsqps{{\rm\thinspace cm^2 s^{-1}}}

\def\K{{\rm\thinspace K}}
\usepackage{wasysym}

\title[Tracing gas motions in the Centaurus Cluster]{Tracing gas motions in the Centaurus cluster}
\author[J. Graham, A.C. Fabian, J.S. Sanders and R.G. Morris]{J. Graham$^1$\thanks{E-mail:jgraham@ast.cam.ac.uk}, A.C. Fabian$^1$, J.S. Sanders$^1$ and R.G. Morris$^{1,2}$\\
\footnotesize$^1$ Institute of Astronomy, Madingley Road, Cambridge\\
$^2$ Kavli Institute for Particle Astrophysics and Cosmology, Stanford
Linear Accelerator Center, Stanford, CA 94305-4060, USA}

\begin{document}

\maketitle

\begin{abstract}
  We apply the stochastic model of iron transport developed by
  \cite{Rebusco2005} to the Centaurus cluster. Using this model, we
  find that an effective diffusion coefficient $D$ in the range
  $2\times10^{28} - 4\times10^{28}\cmsqps$ can approximately reproduce
  the observed abundance distribution. Reproducing the flat central
  profile and sharp drop around $30-70\kpc$, however, requires a
  diffusion coefficient that drops rapidly with radius so that $D >
  4\times10^{28} \cmsqps$ only inside about $25 \kpc$. Assuming that
  all transport is due to fully-developed turbulence, which is also
  responsible for offsetting cooling in the cluster core, we calculate
  the length and velocity scales of energy injection. These length
  scales are found to be up to a factor of $\sim 10$ larger than
  expected if the turbulence is due to the inflation and rising of a
  bubble. We also calculate the turbulent thermal conductivity and
  find it is unlikely to be significant in preventing cooling.
\end{abstract}


\section{Introduction}

Galaxy clusters contain bright, extended, sources of X-ray emission.
This emission is primarily thermal bremsstrahlung in the low-density
($\sim 10^{-3} \cm^{-3}$), optically-thin plasma of the IntraCluster
Medium (ICM), gravitationally heated to $10^7-10^8 \K$ in the cluster
dark matter potential. Spectroscopic observations have revealed the
ICM to be enriched with metals ejected from the cluster galaxies with
an average background metallicity of about a third solar
\citep{EdgeStewart1991}. Observations show that this value is constant
out to a redshift of $z \sim 1$ \citep{Mushotzky1997, Hashimoto2004}
and a comparison of the abundance distribution with supernova models
has led to the conclusion that this background enrichment is the
result of type II supernova activity early in the cluster's history
\citep[e.g.][]{Finoguenov2000, Tozzi2003}.

Two distinct classes of cluster have been identified based on their
X-ray surface brightness profile. Cool-core clusters are characterised
by a bright peak in their X-ray surface brightness profile, which is
not present in non-cool-core clusters. The cool-core clusters are
believed to correspond to dynamically relaxed systems, i.e. those
which have not undergone a recent merger. The ICM cooling timescale in
these bright central regions is often much shorter than the Hubble
time. It was expected that this short cooling time would result in the
deposition of large amounts of cool gas that would be observed in
these regions --- the cooling flow model \citep[e.g.][]{Fabian1994}.
However, detailed observations of clusters have failed to locate gas
below about a third of the mean cluster temperature
\citep[e.g.][]{Allen2001, Peterson2003, Sanders2004, Voigt2004} and it
is now widely believed that some form of distributed heating must be
taking place.

A wide range of solutions to the cooling flow problem have been
proposed \cite[see][ for a review]{PetersonFabian2005}, with most of
the attention focused on energy injection by the central Active
Galactic Nucleus (AGN). These models have the advantage that the AGN
is able to provide enough power to offset the cooling
\citep{TaborBinney1993, Churazov2002}; however, it is unclear how the
energy can be distributed to provide enough heating over the entire
cooling region. A clue may be provided by the observation of bubbles
of synchrotron-emitting plasma in many cool-core clusters
\citep[e.g.][]{Bohringer1993, Sanders2002, McNamara2000, Birzan2004}.
These are inflated by jets emanating from the central AGN. Some
clusters show evidence for further ``ghost'' bubbles at larger radii
\citep[e.g.][]{Fabian2002}. These show weak, low-frequency, radio
emission and are presumably the result of an earlier epoch of AGN
activity. This has led to much work on the possibility that the
inflation and subsequent rise of these bubbles through the cluster
atmosphere is responsible for the transfer of energy from AGN to ICM
\citep[e.g.][]{Fabian2003,Mathews2003, Ruszkowski2004}.

There is speculation that cluster heating might be driven by turbulent
motion of the cluster gas. Evidence for a Kolmogorov-type turbulent
cascade has been found in the pressure fluctuation spectrum of Coma
\citep{Schuecker2004} and in the Faraday rotation power spectrum of
Hydra A \citep{VogtEnsslin2005}. However, the level of turbulence in
cool-core clusters remains unclear; the observation of extended,
linear H$\alpha$ filaments in several cool-core clusters has been used
as evidence that the ICM in these systems cannot be fully turbulent
\citep{FabianHalpha}.

A method of quantifying the level of stochastic motions in cool-core
clusters was developed and applied to the Perseus cluster by
\cite{Rebusco2005}. Their method is based on the central metallicity
peak observed in cool-core clusters which is absent in non-cool-core
systems \citep{Fukazawa2004, deGrandi2004}. This is likely a result of
metal ejection by the Brightest Cluster Galaxy (BCG). The abundance
profile does peak on the BCG, but does not match the light profile in
detail. By assuming the difference in these profiles to be the result
of diffusion of metals caused by stochastic motions in the cluster
gas, \citet{Rebusco2005} were able to determine the diffusion
coefficient; $D=2\times10^{29}\cmsqps$ for the Perseus cluster.
Linking this with a model of turbulent heating, and assuming heating
balanced cooling, they predicted the length and velocity scales of the
gas motion as $l\sim 10 \kpc$ and $v\sim 300 \km \s^{-1}$.

Here we apply the model of \cite{Rebusco2005} to the case of the
Centaurus cluster. Centaurus is unusual in showing super-solar
abundances near the centre - the peak metallicity is around 2 compared
with about 0.7 for Perseus - making it an interesting limiting case
for the model.

We adopt the cosmology $H_0=70 \km \s^{-1} \mpc^{-1}$, $\Omega_m =
0.3$, $\Omega_{\Lambda} = 0.7$ and use the solar abundances of
\cite{AndersSolar1989} throughout.

\section{Diffusion Model}

We employed the model for iron transport developed in
\cite{Rebusco2005}. This is based on a combined diffusion-advection
equation in which the advection term ensures that the equations
satisfy the physical constraint that a spatially uniform abundance
undergoes no evolution in the absence of sources:

\begin{equation}
  \frac{\partial n_{H}a}{\partial t} = \nabla \cdot( n_H D\nabla(a)) + S,
\end{equation}
$n_H$ is the ICM hydrogen density, $a$ is the abundance, $D$ is the
diffusion coefficient and $S$ the (spatially distributed) sources of
iron.

The model of iron injection is taken from \cite{Bohringer2004} and
considers two sources of iron; type Ia supernovae and stellar winds.
In each case, it is assumed that the rate of injection was greater in
the past when the stellar population of the BCG was younger and that
the change in time can be approximated using the power law model of
\cite{Renzini1993} leading to source terms of the form:

\begin{equation}
S_{\text{SN}} =  0.7 \times 10^{-12} \times SR \times \left (\frac{t}{t_h} \right )^{-k} \left ( \frac{L_{B}(r)}{L_{B \astrosun}}\right )\thinspace \text{M}_{\astrosun} \text{yr}^{-1},
\end{equation}

\begin{equation}
S_{\text{Stellar}} = 7\times10^{-14}  \left (\frac{t}{t_h} \right)^{-1.3} \left ( \frac{L_B(r)}{L_{B \astrosun}}\right )\thinspace \text{M}_{\astrosun} \text{yr}^{-1}.
\end{equation}

Here $L_{B \astrosun}$ and $\text{M}_{\astrosun}$ refer to the solar
{\it B}-band luminosity and mass, respectively, $t_h$ is the Hubble time, $SR$ is
the supernova rate in SNU (one SNU is the rate of supernovae
explosions such that a galaxy with $10^{10} L_{B \astrosun}$ has one
SN per century), $k$ controls the rate of type IA supernovae in the
past and $L_B$ describes the luminosity profile of the galaxy. The
model parameters $k$, $SR$ and the total integration time are not
known. We expect $k$ in the range $1.1-2$ \citep{Bohringer2004}. The
integration time is interpreted as the time since the last major
cluster merger, as such a process would have a disruptive effect on
the abundance peak. This time together with the values of $k$ and $SR$
are subject to the constraint that the total mass injected match the
observed excess in the cluster. Following \cite{Rebusco2005}, we take
$k=1.4$, integrate for $8\gyr$ and fix the supernova rate at 0.19 SNU
to match the observed mass excess.

In addition to iron injection, there will be evolution of the ICM
hydrogen density with time; for example, stellar mass-loss liberates
hydrogen as well as heavy elements, and merging subsystems will have
gas stripped. We neglect these effects and regard the hydrogen density
as a constant in all our models.

\section{Modelling diffusion in Centaurus}

\subsection{Properties of the Centaurus Cluster}

High resolution X-ray imaging of the core of the Centaurus cluster has
revealed a complex, asymmetric, structure with a wealth of small-scale
features \citep{FabianCentaurusDeep}. For our purposes it is
sufficient to consider a spherically symmetric profile. In order to
construct an accurate profile over a wide range of cluster radii, we
have combined data from several observations of Centaurus. The highest
resolution \chandra\ data is only available for the inner $\sim
60\kpc$ of the cluster and so we have supplemented this with a new
reduction of archival \xmm\ data out to $\sim100\kpc$ and published
deprojections of data from the \rosat\ \citep{Allen1994} and \asca\
\citep{White2000} satellites at larger radii.

\subsubsection{Data Reduction}

We analysed \chandra\ observation IDs 0504, 0505, 4954, 4955 and
5310. Each observation placed the centre of the cluster near the
aim-point of the ACIS-S3 CCD. The dataset was reprocessed using
\textsc{ciao} and the latest appropriate gain file
(acisD2000-01-29gain\_ctiN0003.fits). Flares in the datasets were
filtered by using the $2.5-7\kev$ count rate on the ACIS-S1 CCD
(which is back-illuminated like the S3 CCD), and the
\textsc{lc\_clean} tool. This yielded a total exposure time of $196\ks$.
The event files were reprojected to match the longest observation
(4954). Standard blank-sky observations were tailored to match the
individual observations, and then reprojected to match 4954. The
exposure time in each background observation was adjusted so that the
count rate in the $9-12\kev$ band matched the cluster observation, to
account for the gradual increase in background with time.

In each annulus, spectra and background spectra were extracted from
each of the datasets and blank-sky observations. Weighted response
matrices were generated for each of the different observations using
\textsc{mkacisrmf}, and weighted ancillary responses created using
\textsc{mkwarf}. The extracted spectra in the annulus were added
together. We took the background spectra for each of the datasets. The
exposure time of the spectra were effectively reduced by discarding
photons. The exposure times were reduced so that the fractional time
of each background observation to the total exposure time of the
background spectrum matched the fractional time of each observation to
the total exposure time of the total foreground spectrum. This
procedure also optimised for the longest total background exposure
time. The shortened background spectra were added together to create a
total background. The responses were added together, with weighting
according to the number of counts as before.

The spectra were fit using the \textsc{projct} model in \textsc{xspec}
\cite{Arnaud1996} to account for projection. A \textsc{vmekal}
\citep{Mekal1985, Mekal1995} model was used to compute the emission
spectra. In the fit the Galactic absorption was a free parameter. In
each shell, the temperature, normalisation, O, Ne, Si, S, Ar, Ca, Fe
and Ni abundances were free parameters. The model was fit between
$0.5$ and $7\kev$, minimising the $\chi^2$ statistic.

To obtain the 1D profile from the \chandra\ observations, we averaged over a 93.9-degree sector to the west of the centre, so avoiding the swirl of cool gas seen to the east \citep{FabianCentaurusDeep}.

The \xmm\ data were reduced in standard fashion. High-energy ($10-15
\kev$), single pixel ({\tt \hbox{PATTERN == 0}}) light curves were
constructed in 100-second bins for the events satisfying the FLAGs
{\tt \#XMMEA\_EM} / {\tt \#XMMEA\_EP} for the MOS / pn, respectively.
Count-rate cuts of $0.15 / 0.45 \ctps$ were applied for the two
instruments. This results in $31\ks$ of good time for the MOS, but
zero for the pn due to background flaring.

The region between $0-150\kpc$ was divided into 10 annuli with equal
net counts. Background spectra were extracted from the blank-sky
fields of \cite{ReadPonman2003}, and normalised using the ratio of the
count-rate outside the field of view for the science to background
observations. The results presented are from MOS1 only. The model was
{\tt \hbox{projct*phabs(mekal)}} with a common, free abundance, and
spectra were fitted over the range $0.5-7.5 \kev$.

\subsubsection{Adopted profiles}

\begin{figure}
  \includegraphics[width=7cm]{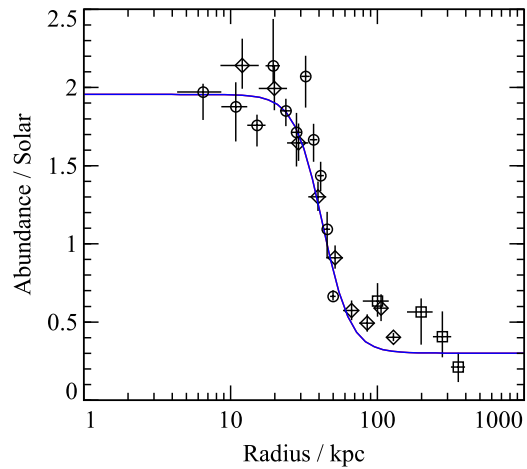}
  \caption{Deprojected iron abundance profile for the Centaurus cluster. The data is from \chandra\ (circles), \xmm\ (diamonds) and \rosat\ (squares). \chandra\ data for the inner $5\kpc$ (where an abundance hole is observed) has been excluded. The solid line represents a fitted profile with a background abundance of 0.3 solar, adopted through the remainder of the work.}
\label{fig:abundances}
\end{figure}

The deprojected iron abundance profile used is shown in Fig.
\ref{fig:abundances}. We have removed data-points corresponding to a
central abundance decrease. Such an ``abundance hole'' could be due to
the gas being multiphase. If it is real, it will certainly require
some additional physical process beyond diffusion from a centrally
concentrated source, e.g. the radial gas motion associated with bubble
formation \citep{Churazov2004}. We have fitted a model profile of the
form:
\begin{equation}
   a = 0.3 \frac{19.2+\left ( \frac{r/\kpc}{33.4}\right )^{4.55}}{2.95+\left (\frac{r/\kpc}{33.4} \right )^{4.55}}a_{\astrosun}.
\end{equation} 

\begin{figure}
  \includegraphics[width=7cm]{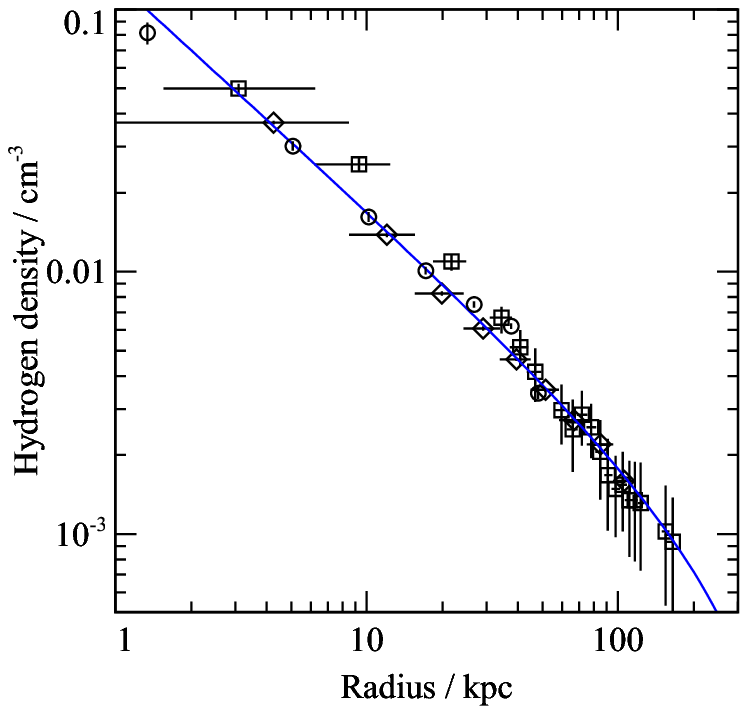}
  \caption{Deprojected hydrogen density with data from \chandra\ (circles), \xmm\ (diamonds) and \rosat\ (squares). The solid line is a power-law fit}.
\label{fig:hydrogen}
\end{figure}

\begin{figure}
  \includegraphics[width=7cm]{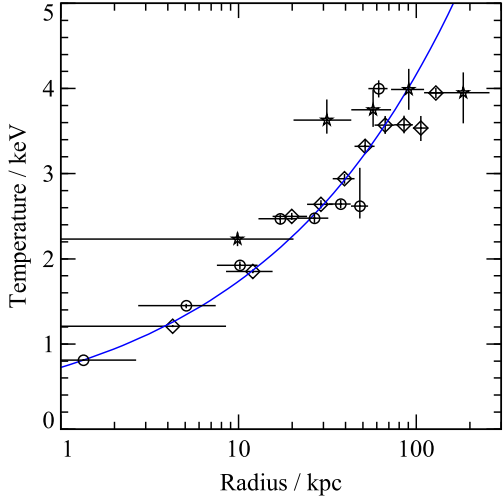}
  \caption{Deprojected temperature profile for the Centaurus cluster with data from \chandra\ (circles), \xmm\ (diamonds) and \asca\ (stars) The solid line is a simple power-law fit to the profile. This form provides a good fit in the inner 100\kpc\ of the cluster but may not be accurate outside this region.}
\label{fig:temperature}
\end{figure}

The deprojected hydrogen density profile is shown in Fig. \ref{fig:hydrogen} with the adopted fit of the form:

\begin{equation}
  n_{H} = \left (\frac{r/\kpc}{0.096} \right )^{-0.87} - 0.00055 \cm^{-3} .
\end{equation}
It has been assumed that the electron and hydrogen densities are related as $n_e = 1.2n_H$. 
The deprojected temperature profile is shown in Fig. \ref{fig:temperature}, with a powerlaw fit of the form:
\begin{equation}
  T = 0.72\left ( \frac{r}{\kpc}\right )^{0.38} \kev ,
\end{equation}
which is valid over the range $1<r<100\kpc$.

\subsection{Properties of NGC\,4696}

NGC\,4696 is the brightest cluster galaxy of the Centaurus cluster and
is situated at the centre of the abundance peak. We derive the
luminosity of NGC\,4696 from the integrated photometry presented in
the HYPERLEDA catalogue \citep{HYPERLEDA}. The total B-band magnitude
is given as $\text{B}=11.51\pm0.01$, uncorrected for absorption, and
$\text{B}=11.03$ with the stated correction for galactic B-band
extinction. Hence we adopt a {\it B}-band luminosity for NGC\,4696 of
$1.2\times 10^{11} L_{\odot}$. The luminosity profile was taken to be
a \citet{Hernquist1990} profile with a scale radius of $9.9 \kpc$
chosen to match the radius containing 50 per cent of the total {\it B}
light in \cite{ESOgalaxies}.

\section{Results}
\begin{figure}
\includegraphics{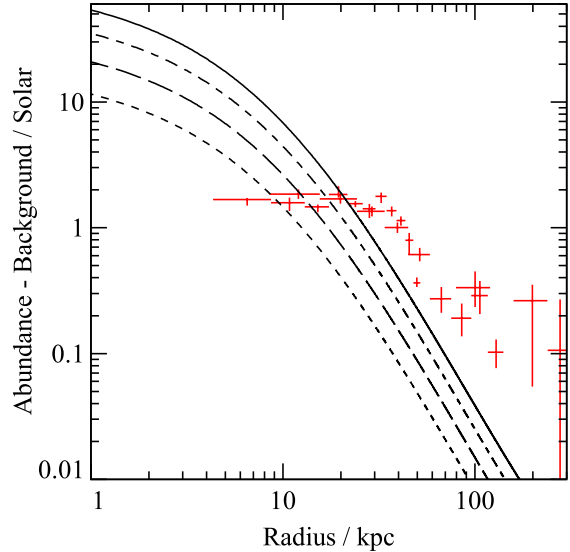}
\caption{Background-subtracted abundance profile in the absence of any diffusion; i.e. when the abundance follows the BCG light. Points represent the data, as in figure \ref{fig:abundances}, whilst lines represent the model at times of 1,2,4 and 8 \gyr.}
\label{fig:D0}
\end{figure}

\begin{figure*}
\includegraphics[width=\textwidth]{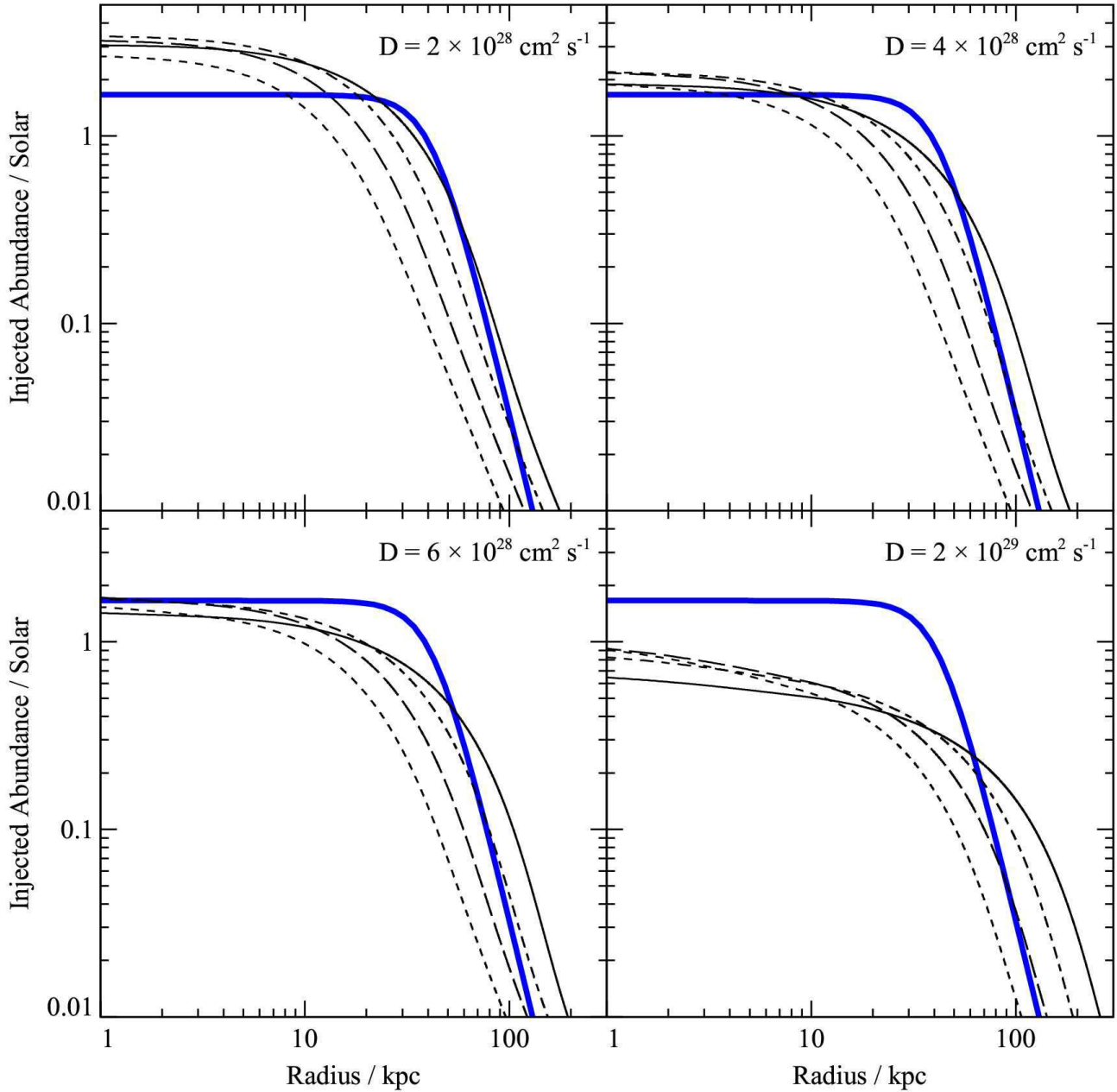}
\caption{Simulations with constant diffusion coefficients of $2 \times 10^{27} \cmsqps$, $2 \times 10^{28} \cmsqps$, $4 \times 10^{28} \cmsqps$ and $2 \times 10^{29} \cmsqps$ at times of 1, 2, 4 and 8 \gyr\ (dotted, dashed, dot-dashed and solid lines respectively) and background-subtracted model (thick solid line).}
\label{fig:results_static_D}
\end{figure*}

Fig. \ref{fig:results_static_D} shows the abundance profile produced
by simulations using spatially uniform diffusion coefficients of $2
\times 10^{28} \cmsqps$, $4 \times 10^{28} \cmsqps$ and $2 \times
10^{29} \cmsqps$. For comparison, Fig. \ref{fig:D0} shows the
abundance profile in the absence of diffusion. $D = 4 \times
10^{28}\cmsqps$ provides a reasonable fit to the observed abundance
profile in the central regions, although the model significantly
underestimates the iron abundance at higher radii, whilst $D = 2
\times 10^{28}\cmsqps$ provides a better fit at outer radii but is a
poor fit within 20\kpc\ of the core. These values of $D$ are factors 5
and 10 times lower than the value of $D=2 \times 10^{29} \cmsqps$
found for Perseus by \cite{Rebusco2005} and are consistent with the
limit $D < 6 \times 10^{28} \cmsqps$ derived by
\cite{FabianCentaurusDeep} based on the width of the abundance peak
and the local sound crossing time.

\subsection{Variable diffusion coefficient}\label{sec:variable_D}

\cite{Rebusco2005} find that the abundance profile becomes more
``boxy'' when the diffusion coefficient declines as a function of
radius. This can be understood, as the large diffusion coefficient
near the cluster centre causes metals to be rapidly spread to larger
radii where they accumulate. For the Perseus cluster, there is no
clear evidence that such a model reproduces the observed abundance
distribution. In Centaurus, however, the high central abundance region
is more sharply cut off, indicating that such a model might be
favoured. Here, we consider fiducial models of the form proposed by
\cite{Rebusco2005}:

\begin{equation}
  D = D_{0} \left( \frac{{n_H}(r)}{{n_H}(r_0)} \right) ^ \alpha ,
\end{equation}
with $r_0$ being a characteristic scale radius at which the diffusion
coefficient is $D_0$, and $\alpha$ controlling how centrally
concentrated the diffusion coefficient is (larger values of $\alpha$
correspond to more centrally concentrated diffusion coefficients).
However, such models may lead to unphysical values of the diffusion
coefficient in some areas of the cluster and so we impose the
additional constraint that $D \leq 0.11 c_{s} r$ where $c_s$ is the
local sound speed. The numerical factor is taken from the turbulent
diffusion simulations of \cite{DennisChandran2004} that are discussed
in Section \ref{sec:dissipation}.

\begin{figure}
\includegraphics{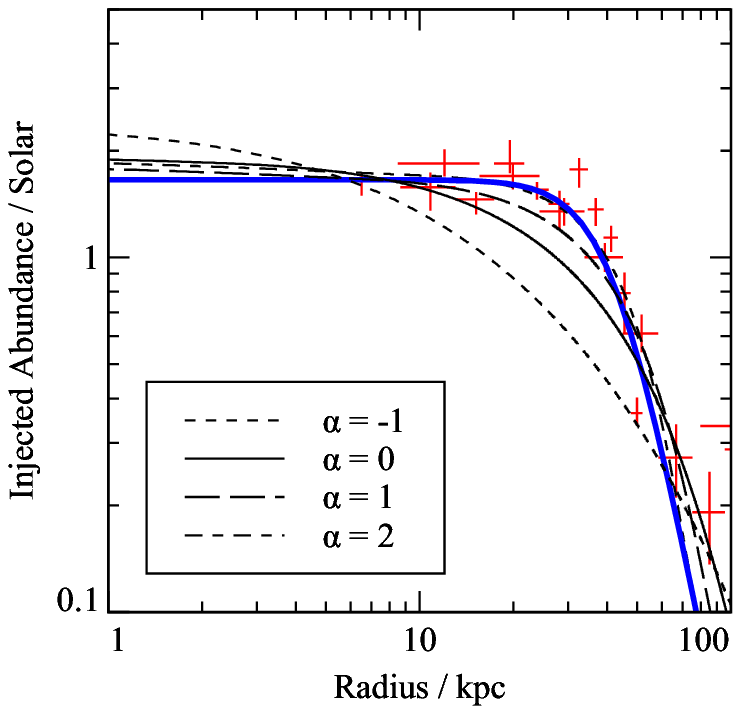}
\caption{Variable diffusion coefficient model at an age of $8 \gyr$ with $D = 4\times10^{28} \cmsqps \frac{n(r)}{n(r_0)}^{\alpha}$, $r_0= 25 \kpc$, $\alpha=-1$ (dots), $\alpha=0$ (solid) $\alpha=1$ (dashed) $\alpha=2$ (dot-dashed) and background-subtracted model (thick solid line). Models with increasing values of $\alpha$ provide an improved fit to the data points.}
\label{fig:results_variable_D}
\end{figure}

Fig. \ref{fig:results_variable_D} shows abundance profiles with values
of $\alpha$ from -1 to 3 and a scale radius of $25 \kpc$. It can be
seen that the flat inner profile and sharp curvature around $30 - 40
\kpc$ is better matched with increasing values of $\alpha$. This is
consistent with a picture where the energy driving the turbulent
motion is injected at the cluster centre, possibly by the inflation
and subsequent buoyant rise of the radio bubbles.

\section{Turbulent heating of cluster gas}
The dissipation of energy contained in turbulent motions of the
cluster gas may act as a significant heat source in the ICM. This
process can provide the kind of spatially extended source of heat that
is needed to balance cooling over the entire cooling radius of the
cluster. There are two distinct sources for such energy; the merging
of sub-clusters and energy injection through the central AGN. These
processes differ in the expected length scales of the energy
injection. In the case of mergers, this scale can be comparable to the
size of the cluster, whilst in the case of AGN activity, the scale
will be much smaller, limited to the size of the radio bubbles.
Turbulence may also produce heating by mixing of hotter gas into the
cooler regions, i.e. turbulent conduction \citep[e.g.][]{Cho2003,
  Voigt2004}.

\subsection{Heating by dissipation}\label{sec:dissipation}

On dimensional grounds, a diffusion coefficient may be written as $D =
C_1 v l$ where $v$ and $l$ are velocity and length scales relevant to
the diffusion process and $C_1$ is a constant, typically of order
unity. By a similar argument, the simplest possible expression for the
rate of heating due to dissipation turbulent motions is $\Gamma = C_2
\rho v^3 / l$, where $\rho$ is the gas density and $v$ and $l$ are the
energy injection scales of the turbulence, assumed to be the same as
the scales of the diffusive motions. If we adopt these expressions and
assume the cluster cooling to be entirely balanced by turbulent
dissipation, we may write:

\begin{equation}\label{eqn:turbulent_heating}
  \Lambda[T(r),a(r)]n(r)^{2} = \frac{C_2 \rho(r) v(r)^3}{l(r)},
\end{equation}
where $\Lambda(T,Z)$ is the cooling function. To fix the constants
$C_1$ and $C_2$ we take the values given in \cite{DennisChandran2004};
$C_{1}=0.11$ and $C_{2} = 0.4$. In practise, Equation
\ref{eqn:turbulent_heating} is an upper limit on the required
turbulent heating, as any process that induces turbulent motion will
also provide other heating. Observations show both weak shocks and
sound waves associated with rising bubbles which will provide
additional distributed heating.

\begin{figure*}
\includegraphics{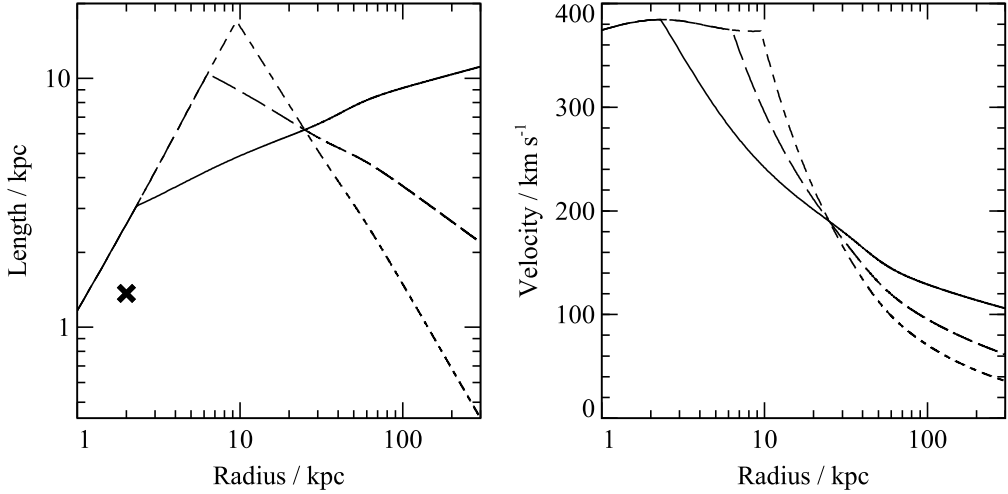}
\caption{Length and velocity scales needed for turbulent diffusion to balance heating with in a model with $D_0=4 \times 10^{28} \cmsqps$, $r_0 = 25\kpc$ and $\alpha=0$ (solid line), $\alpha=1$ (dashed line) and $\alpha=2$ (dot-dashed line). The cross indicates the position and average radius of the Centaurus bubbles \citep{Dunn2004}.}
\label{fig:length_velocities}
\end{figure*}

Fig. \ref{fig:length_velocities} shows the calculated length and velocity scales of the turbulent motions required to balance cooling, given the variation in D with radius. The cooling function was calculated using the {\sc Mekal} plasma emission code \citep{Mekal1985, Mekal1995}, assuming that the ratio of all heavy element abundances to solar follow that of iron. At the scale radius, the models have a scale length of about $4\kpc$ and a velocity of about $150\kmps$. The models with $\alpha$ larger than 0 show an initial rise in the length scale of turbulent motions --- this is over the region where the diffusion coefficient is limited by the requirement that $D \leq 0.11rc_s$ --- and is the region in which the diffusion approximation is least reliable. Outside this inner region, the rapid drop-off in both length and velocity scales is again consistent with a picture of turbulent energy being injected into the cluster core and dissipating away from this region.

\begin{figure*}
\includegraphics{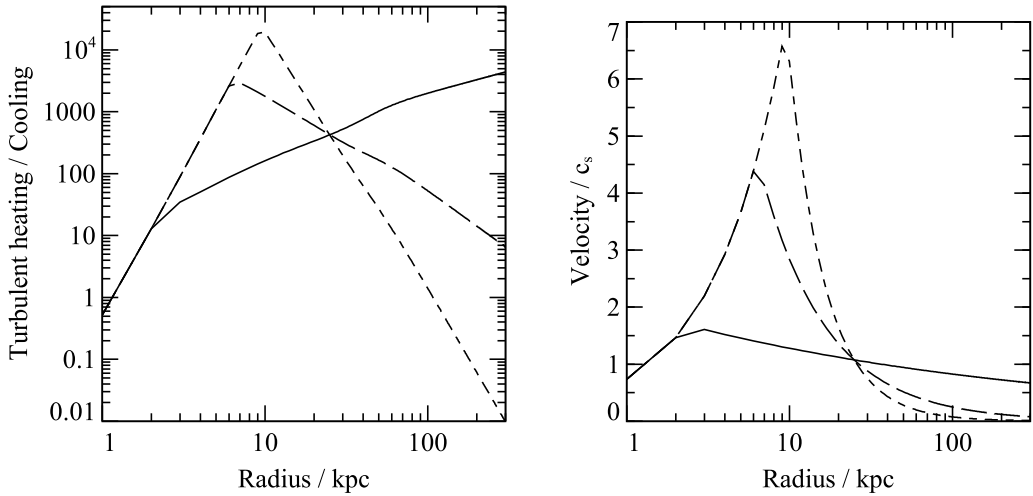}
\caption{Heating rates as a fraction of the required heating and the velocity as a fraction of local sound speed, assuming $D = 0.11vl_{\text{max}}$ (see text for definitions) in a model with $D_0=4 \times 10^{28} \cmsqps$, $r_0 = 25\kpc$ and $\alpha=0$ (solid line), $\alpha=1$ (dashed line) and $\alpha=2$ (dot-dashed line).}
\label{fig:turbulent_heating_fixed_l}
\end{figure*}

\begin{figure*}
\includegraphics{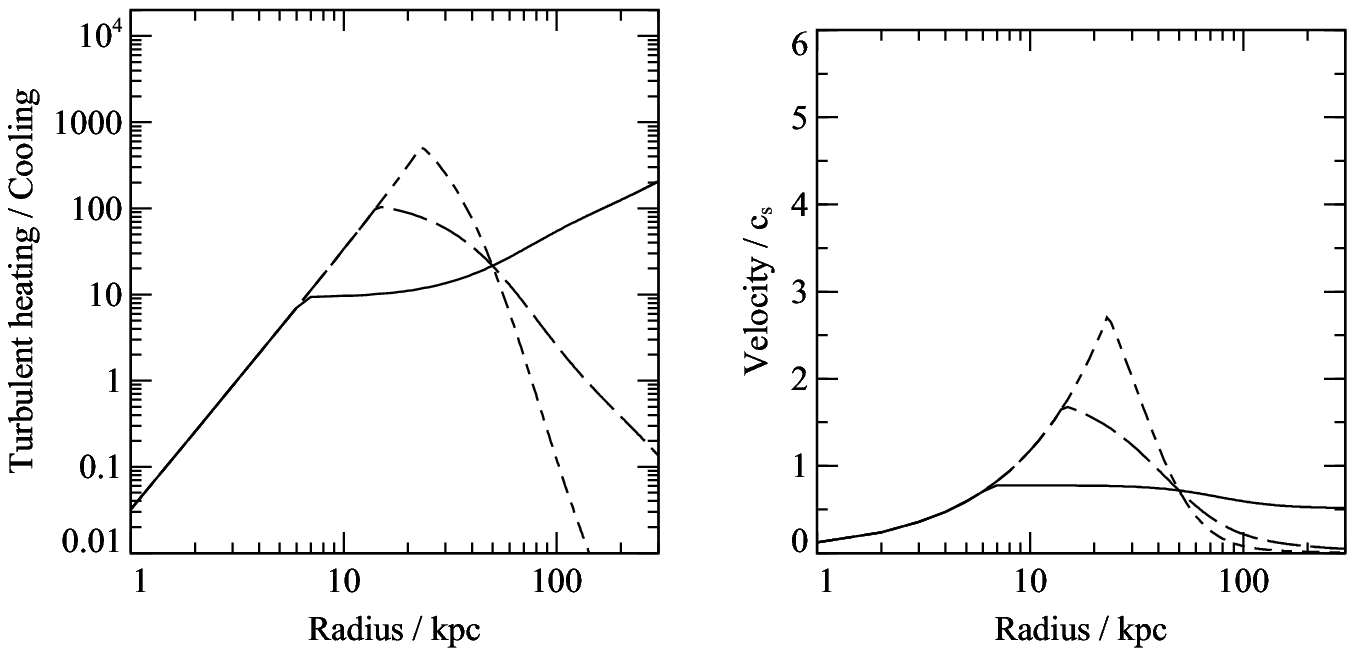}
\caption{As Fig. \ref{fig:turbulent_heating_fixed_l} but for the Perseus cluster using the results of \protect \cite{Rebusco2005}; $D_0=2 \times 10^{29} \cmsqps$, $r_0 = 50\kpc$ and $\alpha=0$ (solid line), $\alpha=1$ (dashed line) and $\alpha=2$ (dot-dashed line).}
\label{fig:turbulent_heating_fixed_l_perseus}
\end{figure*}

Using the calculated length and velocity scales, we may make a
comparison of the implied turbulent viscosity with the standard
\cite{Spitzer1956} viscosity for a hydrogen plasma, assuming that the
appropriate turbulent viscosity is of the form $\mu = \alpha \rho v
l$. Using $vl = D/0.11$ and the cluster properties at a radius of
25\kpc, the Spitzer viscosity is $2.5\times 10^2 \g \cm^{-1} \s^{-1}$,
and the turbulent viscosity $\mu = 4.5\times 10^3\alpha$, so the
turbulent viscosity exceeds the ion viscosity for any $\alpha > 0.06$.

It is notable that, in all the models, the length scale of the
turbulence rises a factor of more than 3 above the mean radius
measurement for the Centaurus bubbles of $1.37 \kpc$ from
\cite{Dunn2004}. If we consider this bubble radius as the upper bound
on the energy injection radius $l_\text{max}=1.37 \kpc$, then $v(r)
\geq {D(r)}/{0.11 l_\text{max}}$. Fig.
\ref{fig:turbulent_heating_fixed_l} shows the heating as a fraction of
the cooling rate and the velocity as a fraction of the local sound
speed assuming $l=l_\text{max}$ everywhere. The effect on the
calculated heating rate and velocity distribution is shown in Fig.
\ref{fig:turbulent_heating_fixed_l}. Since the heating rate at fixed
$D$ varies as $l^{-4}$, fixing the injection scale of the turbulence
to be small causes large increases in the calculated heat injected.
For models with $\alpha>0$, it also requires that the gas velocity be
strongly supersonic in order to maintain the diffusion coefficient in
the cluster centre. In practise, the supersonic velocities are in a
regime where the simple model of turbulence we have used does not
apply and shock waves would presumably damp such motions quickly.
Nevertheless, such supersonic motions and large heat excesses are not
observed in the cluster core, which suggests that bubble-injected
turbulent motions may not be able to reproduce the observed abundance
distribution.

For comparison, Fig. \ref{fig:turbulent_heating_fixed_l_perseus} shows
the same information for the Perseus cluster, based on the results of
\citet{Rebusco2005}, who did not find any reason to favour the
$\alpha>0$ results over the simple $\alpha=0$ uniform diffusion
coefficient. The value of $l_\text{max}$ is fixed at $8.52 \kpc$ based
on the average inner bubble radius from \citet{Dunn2004}. In the case
of Perseus, the model does not produce supersonic velocities, at least
for the case of a uniform diffusion coefficient. Application of the
technique to a larger sample of clusters would be needed to show
whether the problems seen in Centaurus are a special feature of that
cluster.

It is clear that reproducing the observed abundance distribution with
a diffusion model requires both a large diffusion coefficient at the
centre and a relatively small coefficient ($D \lesssim 4 \times
10^{28} \cmsqps$) at radii of about $25\kpc$ to prevent spreading of
the sharp abundance edge. On the assumption that turbulence is
produced by the rising bubbles, this situation is hard to reconcile
with the presently observed bubble properties. This may be because the
presently observed bubbles are not representative of those produced
during the lifetime of the cluster; if previous generations of bubbles
have been substantially larger, the diffusion coefficient can be
maintained with lower gas velocities. However the power required to
inflate the bubbles scales as $r_{\text{bubble}}^2$ and so the
increase in bubble radius required to bring the diffusive velocity
under the sound speed under the assumptions used in Fig.
\ref{fig:turbulent_heating_fixed_l} would require more than an order
of magnitude increase in jet luminosity in the models with $\alpha>0$.

The difficulties may also be avoided if the turbulence is produced by
a different mechanism, e.g. focusing of sound waves produced in
sub-cluster mergers \citep{Pringle1989}. Centaurus is believed to be
undergoing a merger \citep{Churazov1999}. If this merger is affecting
the abundance distribution, for example by causing the sharp
truncation of a naturally extended profile, comparison of the
diffusive models to the observed distribution will be misleading.

If the merger has not affected the abundance distribution, we must
consider the possibility that some process other than turbulence has
been responsible for the observed abundance
distribution. \cite{Crawford2005} observed spatially coincident
H$\alpha$ and soft X-ray filaments surrounding the BCG and conclude
that the structure of the filaments is inconsistent with turbulence at
scales above a few $\kpc$. Similar filaments are observed in the core
of the Perseus cluster \citep{Conselice2001} and spectroscopic
measurements of their line-of-sight velocity (Hatch et. al. 2005,
submitted) indicates that the velocity gradients of the filaments are
smooth, which is a strong indication that the surrounding medium is
not turbulent. Velocity measurements of  a horseshoe-shaped filament
located below a rising bubble are consistent with laminar flow induced
by the bubble rising in a viscous fluid, as proposed by
\cite{FabianHalpha} based on the filament shape. These observations
suggest filaments may be dragged out of the cluster centre by the
rising bubbles, a process that has also been implicated in reproducing
observed features in the Perseus abundance map
\citep{Sanders2005}. Since these motions are largely radial, they
should be more efficient at redistributing the metals than turbulent
motion and we can estimate a diffusion coefficient $D\sim
v_\text{bubble}r_\text{bubble}$. Using the radii and buoyancy
velocities from \cite{Dunn2004} as an estimate of $r_\text{bubble}$
and $v_\text{bubble}$, we find a diffusion coefficient of $D\sim2
\times 10^{29} \cmsqps$, which will be reduced by the fact that bubble
related flows operate in only a fraction of the cluster core at a
time. Spectroscopic measurements of the filaments in Centaurus are
needed to confirm the flow pattern is similar to that in the Perseus
cluster. 

A second possibility is suggested by Fig. \ref{fig:D0}. It is clear
that the the largest discrepancy between the observed abundance
distribution and the naive abundance-following-light model occurs in
the highly enriched region inside about 20\kpc. The metal injection
into these regions would enhance the cooling rate in these regions,
potentially causing the gas to cool out of the X-ray band. The mass of
this gas is of the order $10^8 M_{\astrosun}$, therefore this would not
represent a substantial cooling rate over the lifetime of the
cluster. 

These results suggest the diffusion model may be systematically
overestimating the level of turbulence. It should, however, be
emphasised that the models of turbulent heating and diffusion are very
coarse and so these results should be regarded as very approximate. In
particular, the result that turbulent motion on the scale size of the
bubble requires velocities close to or above the sound speed to
reproduce the observed iron peak should be regarded as demonstrating
the limitations of the purely stochastic model of metal spreading.

\subsection{Turbulent Conduction}

If we imagine the cluster as being divided into thin shells with each
of the gas properties being constant within the shell, the
conductivity required for heating to balance cooling is given by
\cite{Voigt2004}:

\begin{equation}
  \kappa_{j} = \frac{\sum_{i=1}^{j}\Lambda(T_{i},Z_{i})n_{i}^{2} \Delta V_{i}}{4 \pi r_{j}^{2} \left ( \frac{d T}{d r} \right )_{j}} .
\end{equation}
In the case of an unmagnetised plasma, the theoretical electron conductivity is given by the Spitzer form:

\begin{equation}
  \kappa_{\text{Spitzer}} = 640 \left ( \frac{2 \pi}{m_{e}}\right )^{\frac{1}{2}} \frac{k \epsilon_{0}^{2}}{e^4} \frac{(k T)^{\frac{5}{2}}}{Z \ln \Lambda} ,
\end{equation} 
where $\Lambda$ (not to be confused with the cooling function) is the coulomb logarithm:

\begin{equation}
  \ln \Lambda \approxeq 37.8 + \ln \left [ \ \left ( \frac{T}{10^{8} K} \right )  \left ( \frac{n_e}{10^{-3} \text{cm}^{-3}} \right ) \right ] .
\end{equation}
In practise the conductivity is likely to be suppressed by the presence of magnetic fields which bind electrons to field lines and so greatly reduce the effectiveness of transport perpendicular to the local field direction. The overall degree of suppression is therefore strongly dependent on the field geometry in the cluster centre. In the case of a turbulent magnetic field, \cite{Narayan2001} find that the effective conductivity is about one fifth the Spitzer value.

\begin{figure}
\includegraphics{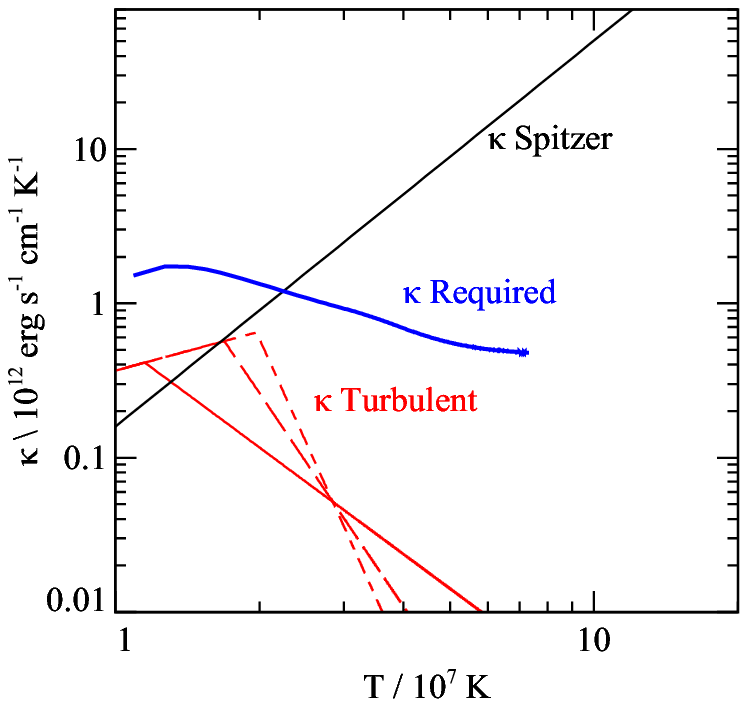}
\caption{Level of turbulent conduction compared to the required conduction for the cluster (solid blue line) and the Spitzer electron conductivity (solid black line). Turbulent diffusion is based on the diffusion model with $D_0=4 \times 10^{28} \cmsqps$, $r_0 = 25\kpc$ and $\alpha=0$ (solid red line), $\alpha=1$ (dashed line) and $\alpha=2$ (dot-dashed line). In all cases, the level of conductivity due to turbulent diffusion is below the conductivity required to balance heating.}

\label{fig:kappa}
\end{figure}

In the case of a turbulent gas, the thermal conductivity $\kappa$ is
given by $\kappa_{\text{turb}} = n_{e} k D$ \cite{Cho2003}. Fig. \ref{fig:kappa} shows the thermal conductivity required to balance conduction in Centaurus with the Spitzer conductivity and the turbulent conductivity derived from the diffusion model in Fig. \ref{fig:results_variable_D}. The turbulent conductivity is lower than the required conductivity everywhere in the cluster. However, it appears that Spitzer electron conductivity may be significant in the outer regions of the cluster, although this depends on the degree of suppression due to the magnetic field. 

\section{Conclusions}

The observed central abundance peak in the Centaurus cluster can be approximately reproduced by diffusive spreading of metals ejected from the BCG. The boxy abundance profile favours a model where the diffusion coefficient declines away from the centre; this is qualitatively consistent with the gas motions resulting from the rising radio bubbles observed in the cluster core. In particular, reproducing the sharp observed abundance drop requires a diffusion coefficient of $D<4\times10^{28}\cmsqps$ for $r>25\kpc$ and hence the velocity field must be relatively quiet in this region. This is a robust result and our main conclusion.

We have calculated approximate length and velocity scales for turbulent motion to heat the cluster. They predict length scales of up to $\sim 10 \kpc$ and velocities of up to $\sim 400 \kmps$ in the inner $\sim 25\kpc$. That this length scale is larger than the bubble scale is a remaining difficulty as a substantial reduction in the length scale of the motion implies a reduction of the diffusion coefficient, producing a poorer fit to the data, or a large increase in the turbulent velocity. This suggests that the level of turbulence may not be as high as predicted. Spreading of metals in the region $r<25\kpc$ may be due to some other process such as the velocity flow behind rising bubbles. More work is needed to quantify how effective such a mechanism can be. With the next generation of X-ray spectrometers, direct measurements of turbulent velocities will be possible and will place strong constraints on the model.

 The thermal conductivity arising from turbulent mixing has also been calculated and this has been shown to be insignificant except, perhaps, in the very central regions of the cluster.

\section*{Acknowledgements}
JG acknowledges support from PPARC and RGM thanks PPARC for support whilst at Cambridge. ACF thanks the Royal Society for support. Work supported in part by the U.S. Department of Energy under contract number DE-AC02-76SF00515.

\bibliographystyle{mnras} 
\bibliography{refs}

\begin{thebibliography}{}

\bibitem[\protect\citeauthoryear{{Allen} \& {Fabian}}{{Allen} \&
  {Fabian}}{1994}]{Allen1994}
{Allen} S.~W.,  {Fabian} A.~C., 1994, \mnras, 269, 409

\bibitem[\protect\citeauthoryear{{Allen}, {Schmidt}, \& {Fabian}}{{Allen}
  et~al.}{2001}]{Allen2001}
{Allen} S.~W., {Schmidt} R.~W.,  {Fabian} A.~C., 2001, \mnras, 328, L37

\bibitem[\protect\citeauthoryear{{Anders} \& {Grevesse}}{{Anders} \&
  {Grevesse}}{1989}]{AndersSolar1989}
{Anders} E.,  {Grevesse} N., 1989, Geochim. Cosmochim. Acta, 53, 197

\bibitem[\protect\citeauthoryear{{Arnaud}}{{Arnaud}}{1996}]{Arnaud1996}
{Arnaud} K.~A., 1996, in ASP Conf. Ser. 101: Astronomical Data Analysis
  Software and Systems V, p.~17

\bibitem[\protect\citeauthoryear{{B{\^i}rzan} et~al.}{{B{\^i}rzan}
  et~al.}{2004}]{Birzan2004}
{B{\^i}rzan} L., {Rafferty} D.~A., {McNamara} B.~R., {Wise} M.~W.,  {Nulsen}
  P.~E.~J., 2004, \apj, 607, 800

\bibitem[\protect\citeauthoryear{{B{\"o}hringer} et~al.}{{B{\"o}hringer}
  et~al.}{2004}]{Bohringer2004}
{B{\"o}hringer} H., {Matsushita} K., {Churazov} E., {Finoguenov} A.,  {Ikebe}
  Y., 2004, \aap, 416, L21

\bibitem[\protect\citeauthoryear{{B{\"o}hringer} et~al.}{{B{\"o}hringer}
  et~al.}{1993}]{Bohringer1993}
{B{\"o}hringer} H., {Voges} W., {Fabian} A.~C., {Edge} A.~C.,  {Neumann} D.~M.,
  1993, \mnras, 264, L25

\bibitem[\protect\citeauthoryear{{Cho} et~al.}{{Cho} et~al.}{2003}]{Cho2003}
{Cho} J., {Lazarian} A., {Honein} A., {Knaepen} B., {Kassinos} S.,  {Moin} P.,
  2003, \apjl, 589, L77

\bibitem[\protect\citeauthoryear{{Churazov} et~al.}{{Churazov}
  et~al.}{2004}]{Churazov2004}
{Churazov} E., {Forman} W., {Jones} C., {Sunyaev} R.,  {B{\"o}hringer} H.,
  2004, \mnras, 347, 29

\bibitem[\protect\citeauthoryear{{Churazov} et~al.}{{Churazov}
  et~al.}{1999}]{Churazov1999}
{Churazov} E., {Gilfanov} M., {Forman} W.,  {Jones} C., 1999, \apj, 520, 105

\bibitem[\protect\citeauthoryear{{Churazov} et~al.}{{Churazov}
  et~al.}{2002}]{Churazov2002}
{Churazov} E., {Sunyaev} R., {Forman} W.,  {B{\"o}hringer} H., 2002, \mnras,
  332, 729

\bibitem[\protect\citeauthoryear{{Conselice}, {Gallagher}, \&
  {Wyse}}{{Conselice} et~al.}{2001}]{Conselice2001}
{Conselice} C.~J., {Gallagher} J.~S.,  {Wyse} R.~F.~G., 2001, \aj, 122, 2281

\bibitem[\protect\citeauthoryear{{Crawford} et~al.}{{Crawford}
  et~al.}{2005}]{Crawford2005}
{Crawford} C.~S., {Hatch} N.~A., {Fabian} A.~C.,  {Sanders} J.~S., 2005,
  \mnras, 363, 216

\bibitem[\protect\citeauthoryear{{De Grandi} et~al.}{{De Grandi}
  et~al.}{2004}]{deGrandi2004}
{De Grandi} S., {Ettori} S., {Longhetti} M.,  {Molendi} S., 2004, \aap, 419, 7

\bibitem[\protect\citeauthoryear{{Dennis} \& {Chandran}}{{Dennis} \&
  {Chandran}}{2005}]{DennisChandran2004}
{Dennis} T.~J.,  {Chandran} B.~D.~G., 2005, \apj, 622, 205

\bibitem[\protect\citeauthoryear{{Dunn} \& {Fabian}}{{Dunn} \&
  {Fabian}}{2004}]{Dunn2004}
{Dunn} R.~J.~H.,  {Fabian} A.~C., 2004, \mnras, 355, 862

\bibitem[\protect\citeauthoryear{{Edge} \& {Stewart}}{{Edge} \&
  {Stewart}}{1991}]{EdgeStewart1991}
{Edge} A.~C.,  {Stewart} G.~C., 1991, \mnras, 252, 414

\bibitem[\protect\citeauthoryear{{Fabian}}{{Fabian}}{1994}]{Fabian1994}
{Fabian} A.~C., 1994, \araa, 32, 277

\bibitem[\protect\citeauthoryear{{Fabian} et~al.}{{Fabian}
  et~al.}{2002}]{Fabian2002}
{Fabian} A.~C., {Celotti} A., {Blundell} K.~M., {Kassim} N.~E.,  {Perley}
  R.~A., 2002, \mnras, 331, 369

\bibitem[\protect\citeauthoryear{{Fabian} et~al.}{{Fabian}
  et~al.}{2003a}]{Fabian2003}
{Fabian} A.~C., {Sanders} J.~S., {Allen} S.~W., {Crawford} C.~S., {Iwasawa} K.,
  {Johnstone} R.~M., {Schmidt} R.~W.,  {Taylor} G.~B., 2003a, \mnras, 344, L43

\bibitem[\protect\citeauthoryear{{Fabian} et~al.}{{Fabian}
  et~al.}{2003b}]{FabianHalpha}
{Fabian} A.~C., {Sanders} J.~S., {Crawford} C.~S., {Conselice} C.~J.,
  {Gallagher} J.~S.,  {Wyse} R.~F.~G., 2003b, \mnras, 344, L48

\bibitem[\protect\citeauthoryear{{Fabian} et~al.}{{Fabian}
  et~al.}{2005}]{FabianCentaurusDeep}
{Fabian} A.~C., {Sanders} J.~S., {Taylor} G.~B.,  {Allen} S.~W., 2005, \mnras,
  360, L20

\bibitem[\protect\citeauthoryear{{Finoguenov}, {David}, \&
  {Ponman}}{{Finoguenov} et~al.}{2000}]{Finoguenov2000}
{Finoguenov} A., {David} L.~P.,  {Ponman} T.~J., 2000, \apj, 544, 188

\bibitem[\protect\citeauthoryear{{Fukazawa}, {Kawano}, \&
  {Kawashima}}{{Fukazawa} et~al.}{2004}]{Fukazawa2004}
{Fukazawa} Y., {Kawano} N.,  {Kawashima} K., 2004, \apjl, 606, L109

\bibitem[\protect\citeauthoryear{{Hashimoto} et~al.}{{Hashimoto}
  et~al.}{2004}]{Hashimoto2004}
{Hashimoto} Y., {Barcons} X., {B{\"o}hringer} H., {Fabian} A.~C., {Hasinger}
  G., {Mainieri} V.,  {Brunner} H., 2004, \aap, 417, 819

\bibitem[\protect\citeauthoryear{{Hernquist}}{{Hernquist}}{1990}]{Hernquist199%
0}
{Hernquist} L., 1990, \apj, 356, 359

\bibitem[\protect\citeauthoryear{{Lauberts} \& {Valentijn}}{{Lauberts} \&
  {Valentijn}}{1989}]{ESOgalaxies}
{Lauberts} A.,  {Valentijn} E.~A., 1989, {The surface photometry catalogue of
  the ESO-Uppsala galaxies}.
\newblock Garching: European Southern Observatory, |c1989

\bibitem[\protect\citeauthoryear{{Liedahl}, {Osterheld}, \&
  {Goldstein}}{{Liedahl} et~al.}{1995}]{Mekal1995}
{Liedahl} D.~A., {Osterheld} A.~L.,  {Goldstein} W.~H., 1995, \apjl, 438, L115

\bibitem[\protect\citeauthoryear{{Mathews} et~al.}{{Mathews}
  et~al.}{2003}]{Mathews2003}
{Mathews} W.~G., {Brighenti} F., {Buote} D.~A.,  {Lewis} A.~D., 2003, \apj,
  596, 159

\bibitem[\protect\citeauthoryear{{McNamara} et~al.}{{McNamara}
  et~al.}{2000}]{McNamara2000}
{McNamara} B.~R. et~al., 2000, \apjl, 534, L135

\bibitem[\protect\citeauthoryear{{Mewe}, {Gronenschild}, \& {van den
  Oord}}{{Mewe} et~al.}{1985}]{Mekal1985}
{Mewe} R., {Gronenschild} E.~H.~B.~M.,  {van den Oord} G.~H.~J., 1985, \aaps,
  62, 197

\bibitem[\protect\citeauthoryear{{Mushotzky} \& {Loewenstein}}{{Mushotzky} \&
  {Loewenstein}}{1997}]{Mushotzky1997}
{Mushotzky} R.~F.,  {Loewenstein} M., 1997, \apjl, 481, L63

\bibitem[\protect\citeauthoryear{{Narayan} \& {Medvedev}}{{Narayan} \&
  {Medvedev}}{2001}]{Narayan2001}
{Narayan} R.,  {Medvedev} M.~V., 2001, \apjl, 562, L129

\bibitem[\protect\citeauthoryear{{Paturel} et~al.}{{Paturel}
  et~al.}{2003}]{HYPERLEDA}
{Paturel} G., {Petit} C., {Prugniel} P., {Theureau} G., {Rousseau} J., {Brouty}
  M., {Dubois} P.,  {Cambresy} L., 2003, VizieR Online Data Catalog, 7237

\bibitem[\protect\citeauthoryear{{Peterson} \& {Fabian}}{{Peterson} \&
  {Fabian}}{2005}]{PetersonFabian2005}
{Peterson} J.~R.,  {Fabian} A.~C., 2005, arXiv:astro-ph/0505517

\bibitem[\protect\citeauthoryear{{Peterson} et~al.}{{Peterson}
  et~al.}{2003}]{Peterson2003}
{Peterson} J.~R., {Kahn} S.~M., {Paerels} F.~B.~S., {Kaastra} J.~S., {Tamura}
  T., {Bleeker} J.~A.~M., {Ferrigno} C.,  {Jernigan} J.~G., 2003, \apj, 590,
  207

\bibitem[\protect\citeauthoryear{{Pringle}}{{Pringle}}{1989}]{Pringle1989}
{Pringle} J.~E., 1989, \mnras, 239, 479

\bibitem[\protect\citeauthoryear{{Read} \& {Ponman}}{{Read} \&
  {Ponman}}{2003}]{ReadPonman2003}
{Read} A.~M.,  {Ponman} T.~J., 2003, \aap, 409, 395

\bibitem[\protect\citeauthoryear{{Rebusco} et~al.}{{Rebusco}
  et~al.}{2005}]{Rebusco2005}
{Rebusco} P., {Churazov} E., {B{\" o}hringer} H.,  {Forman} W., 2005, \mnras,
  359, 1041

\bibitem[\protect\citeauthoryear{{Renzini} et~al.}{{Renzini}
  et~al.}{1993}]{Renzini1993}
{Renzini} A., {Ciotti} L., {D'Ercole} A.,  {Pellegrini} S., 1993, \apj, 419, 52

\bibitem[\protect\citeauthoryear{{Ruszkowski}, {Br{\"u}ggen}, \&
  {Begelman}}{{Ruszkowski} et~al.}{2004}]{Ruszkowski2004}
{Ruszkowski} M., {Br{\"u}ggen} M.,  {Begelman} M.~C., 2004, \apj, 615, 675

\bibitem[\protect\citeauthoryear{{Sanders} \& {Fabian}}{{Sanders} \&
  {Fabian}}{2002}]{Sanders2002}
{Sanders} J.~S.,  {Fabian} A.~C., 2002, \mnras, 331, 273

\bibitem[\protect\citeauthoryear{{Sanders} et~al.}{{Sanders}
  et~al.}{2004}]{Sanders2004}
{Sanders} J.~S., {Fabian} A.~C., {Allen} S.~W.,  {Schmidt} R.~W., 2004, \mnras,
  349, 952

\bibitem[\protect\citeauthoryear{{Sanders}, {Fabian}, \& {Dunn}}{{Sanders}
  et~al.}{2005}]{Sanders2005}
{Sanders} J.~S., {Fabian} A.~C.,  {Dunn} R.~J.~H., 2005, \mnras, 360, 133

\bibitem[\protect\citeauthoryear{{Schuecker} et~al.}{{Schuecker}
  et~al.}{2004}]{Schuecker2004}
{Schuecker} P., {Finoguenov} A., {Miniati} F., {B{\"o}hringer} H.,  {Briel}
  U.~G., 2004, \aap, 426, 387

\bibitem[\protect\citeauthoryear{{Spitzer}}{{Spitzer}}{1956}]{Spitzer1956}
{Spitzer} L., 1956, {Physics of Fully Ionized Gases}.
\newblock Physics of Fully Ionized Gases, New York: Interscience Publishers,
  1956

\bibitem[\protect\citeauthoryear{{Tabor} \& {Binney}}{{Tabor} \&
  {Binney}}{1993}]{TaborBinney1993}
{Tabor} G.,  {Binney} J., 1993, \mnras, 263, 323

\bibitem[\protect\citeauthoryear{{Tozzi} et~al.}{{Tozzi}
  et~al.}{2003}]{Tozzi2003}
{Tozzi} P., {Rosati} P., {Ettori} S., {Borgani} S., {Mainieri} V.,  {Norman}
  C., 2003, \apj, 593, 705

\bibitem[\protect\citeauthoryear{{Vogt} \& {En{\ss}lin}}{{Vogt} \&
  {En{\ss}lin}}{2005}]{VogtEnsslin2005}
{Vogt} C.,  {En{\ss}lin} T.~A., 2005, \aap, 434, 67

\bibitem[\protect\citeauthoryear{{Voigt} \& {Fabian}}{{Voigt} \&
  {Fabian}}{2004}]{Voigt2004}
{Voigt} L.~M.,  {Fabian} A.~C., 2004, \mnras, 347, 1130

\bibitem[\protect\citeauthoryear{{White}}{{White}}{2000}]{White2000}
{White} D.~A., 2000, \mnras, 312, 663

\end{thebibliography}

\end{document}